\documentclass[10pt,journal,compsoc]{IEEEtran}
\IEEEoverridecommandlockouts

\usepackage[dvipsnames]{xcolor}
\usepackage{orcidlink}
\usepackage{cite}
\usepackage{amsmath,amssymb,amsfonts}
\usepackage{algorithm}
\usepackage[noend]{algorithmic}
\usepackage{graphicx}
\usepackage{textcomp}
\usepackage{longtable}
\usepackage{lscape}
\usepackage{lipsum, babel}
\usepackage{amssymb}
\usepackage{pifont}%
\usepackage{tabularray}
\usepackage{enumitem}
\usepackage[newfloat,frozencache,cachedir=.]{minted}
\usepackage{setspace}

\def\BibTeX{{\rm B\kern-.05em{\sc i\kern-.025em b}\kern-.08em
    T\kern-.1667em\lower.7ex\hbox{E}\kern-.125emX}}
\usepackage{lineno}
    
\newif\ifshowmods
\showmodstrue     

\ifshowmods

	\newcommand{\textstrike}[1]{\st{#1}}
\else

	\newcommand{\textstrike}[1]{}
\fi

\begin{document}
%
%
%
%
%
%
%
%
%

\title{BaseSAP: Modular Stealth Address Protocol for Programmable Blockchains}
%
%
%
%

\author{Anton~Wahrstätter\orcidlink{0000-0003-3816-5938}, Matthew~Solomon, Ben~DiFrancesco, Vitalik~Buterin,
          and~Davor~Svetinovic\orcidlink{0000-0002-3020-9556}
\IEEEcompsocitemizethanks{\IEEEcompsocthanksitem A. Wahrstätter is with the Research Institute for Cryptoeconomics, Department of Information Systems and Operations Management, Vienna University of Economics and Business, Vienna, Austria. E-Mail: anton.wahrstaetter@wu.ac.at}%
\IEEEcompsocitemizethanks{\IEEEcompsocthanksitem M. Solomon and B. DiFrancesco are with ScopeLift, Broomall, Pennsylvania, United States. E-Mail: \{matt, ben\}@scopelift.co}%
\IEEEcompsocitemizethanks{\IEEEcompsocthanksitem V. Buterin is with Ethereum Foundation,  Singapore, Singapore. E-Mail: vitalik@ethereum.org}%
\IEEEcompsocitemizethanks{\IEEEcompsocthanksitem D. Svetinovic is with the Research Institute for Cryptoeconomics, Department of Information Systems and Operations Management, Vienna University of Economics and Business, Vienna, Austria, and the Center for Cyber-Physical Systems, Electrical Engineering and Computer Science, Khalifa University, Abu Dhabi, UAE.%
E-mail: dsve@acm.org
}
\thanks{Manuscript received June 2023.}
}

%
%

\markboth{BaseSAP: Modular Stealth Address Protocol for Programmable Blockchains}%
{Wahrstätter \MakeLowercase{\emph{et al.}}}
%



\IEEEtitleabstractindextext{%
\begin{abstract}
Stealth addresses represent an approach to enhancing privacy within public and distributed blockchains, such as Ethereum and Bitcoin. Stealth address protocols generate a distinct, randomly generated address for the recipient, thereby concealing interactions between entities. In this study, we introduce BaseSAP, an autonomous base-layer protocol for embedding stealth addresses within the application layer of programmable blockchains. BaseSAP expands upon previous research to develop a modular protocol for executing unlikable transactions on public blockchains. 
BaseSAP allows for developing additional stealth address layers using different cryptographic algorithms on top of the primary implementation, capitalizing on its modularity. To demonstrate the effectiveness of our proposed protocol, we present simulations of an advanced Secp256k1-based dual-key stealth address protocol. This protocol is designed on top of BaseSAP and is deployed on the Goerli and Sepolia test networks as the first prototype implementation.
Furthermore, we provide cost analyses and underscore potential security ramifications and attack vectors that could affect the privacy of stealth addresses. Our study reveals the flexibility of the BaseSAP protocol and offers insight into the broader implications of stealth address technology.

\end{abstract}

\begin{IEEEkeywords}
Blockchain, Privacy, Security, Confidentiality, Ethereum, Stealth Address 
\end{IEEEkeywords}}

\maketitle
\IEEEdisplaynontitleabstractindextext

%
\IEEEpeerreviewmaketitle


\IEEEraisesectionheading{\section{Introduction}\label{sec:introduction}}


\IEEEPARstart{S}{tealth} addresses have gained importance in blockchain technology due to their potential to improve confidentiality and privacy in transactions on public blockchains. In the context of public blockchains, all transactions are recorded transparently, making it possible to track the transaction history of a particular pseudonymous user. This traceability could occur unintentionally, as the parties involved in a transaction may not have consciously aimed to establish a linkable record. Nonetheless, since blockchains like Bitcoin and Ethereum are transparent and publicly accessible, third parties can analyze the data and potentially identify the participants in a particular transaction.

Stealth address protocols (SAPs) offer a solution to the privacy challenges faced on public blockchains by enabling users to interact confidentially without allowing external observers to link the parties involved in a transaction. At their core, stealth address protocols empower the shielding of recipient information in peer-to-peer (P2P) transactions~\cite{vansaber_2013, todd_2014}.

Stealth addresses can have numerous applications, including but not limited to donations and payroll payments. They can be used in any P2P interaction where privacy concerns demand concealing the connection between two parties. Users can transfer funds using stealth address protocols while protecting the recipient's identity.

The first stealth addresses were initially introduced in the Bitcoin ecosystem and had since undergone continuous refinement. In 2013, Nicolas van Saberhagen described the CryptoNote protocol, which utilized stealth addresses to enhance the privacy of blockchain transactions~\cite{vansaber_2013}. Peter Todd subsequently built upon this concept in 2014 and further improved it~\cite{todd_2014}. Ultimately, stealth addresses were integrated into the Monero blockchain when it launched in 2014~\cite{getmonero}.

In the Ethereum blockchain, stealth addresses can significantly improve confidentiality. They allow users to autonomously generate a unique, one-time address for each transaction instead of relying on a static, publicly identified address. The programmable nature of the Ethereum blockchain facilitates the development of stealth address protocols on top of it, hence leveraging the decentralization and trust attributes of the underlying blockchain~\cite{Buterin2013}.

Despite their inherent promise, stealth addresses in their present form manifest several constraints pertaining to their prospective deployment and efficacy across diverse blockchain frameworks. Regarding privacy-centered blockchains, such as Monero, stealth address protocols have been intrinsically integrated into the fundamental protocol. However, widely-recognized programmable blockchains, such as Ethereum, which do not inherently encompass robust privacy assurances at the foundational protocol layer, necessitate application-layer interventions.~\cite{getmonero}

Within this context, the potential for stealth address protocols on blockchains such as Ethereum becomes evident. They can introduce innovative privacy functions for users, capitalizing on the Turing-complete environment of Ethereum while concurrently harnessing the cohesive modularity characteristic of the base blockchain protocol. This interoperability could also be employed advantageously for Smart Contract wallets, public goods funding, Decentralized Finance (DeFi) systems, or the Non-Fungible Token (NFT) landscape, thereby broadening the potential application areas for these protocols.

Yet, associated scalability obstacles with conventional stealth address protocols must be suitably confronted. These must be resolved to facilitate the mainstream acceptance of stealth addresses while guaranteeing a secure user experience.

To overcome the aforementioned challenges, we develop BaseSAP. BaseSAP is a fully open and reusable stealth address protocol that can reliably offer stealth addresses at the application layer of programmable blockchains such as Ethereum. The protocol aims to provide a lightweight mechanism for users to generate stealth addresses, maintaining complete backward compatibility and requiring no modifications to the core blockchain.
Our proposed base protocol is agnostic to various cryptographic schemes and holds the potential to substantially improve user interactions with stealth addresses in the context of programmable blockchains. BaseSAP comprises a foundational implementation, which includes the reusable functionality required for any trustless stealth address protocol. 

We designed BaseSAP to be fully extendable, thereby enabling the creation of unique stealth address protocols based on particular cryptographic schemes. Examples of such extensions include stealth addresses derived from the Secp256k1 elliptic curve, stealth address protocols based on elliptic curve pairings~\cite{Feng2021, li2022hybrid, Fan_2019} or generated using lattice-based cryptography~\cite{Liu_2019, muhammed2021}. 
The protocol design ensures compatibility and proactively accommodates future quantum-resistant cryptographic schemes that require larger key sizes.

Beyond the base protocol contribution, we create the first practical implementation on top of BaseSAP. We implement an improved dual-key method that relies on the Secp256k1 elliptic curve and employs \emph{view tags} to improve parsing efficiency compared to conventional Dual-Key Stealth Address Protocols (DKSAPs). 
The key contributions of this work are as follows:
\begin{itemize}
\item We present a comprehensive review of stealth addresses research's current status and utilization across diverse blockchains.
\item We identify and address substantial challenges associated with interoperable stealth address protocols, emphasizing privacy concerns and Denial-of-Service (DoS) attack vulnerabilities.
\item We design and develop BaseSAP as a fully open, cohesive, and extendable stealth address protocol to be integrated into Ethereum~\cite{eip5564} in active collaboration with the Ethereum development community.
\item We illustrate the inherent modularity of our protocol, accentuating the significant potential of such approaches when implemented at the application layer of programmable blockchains. This could benefit diverse areas such as Smart Contract wallets, cryptocurrency donation platforms, public goods funding, decentralized finance, and the Non-Fungible Token landscape.
\item We develop a preliminary stealth address prototype that leverages the Secp256k1 elliptic curve and exhibits superior performance in terms of parsing time when compared to existing Stealth Address Protocols~\cite{nerolation2023}.

\end{itemize}

We publish the code base created for this work under an open-source license to ensure reproducibility and transparency~\cite{nerolation2023}. Additionally, we propose the described protocol as an ERC (Ethereum Request for Comment)~\cite{eip5564} to contribute to adopting stealth addresses on programmable and decentralized blockchains.


\section{Related Work}\label{sec:relatedwork}

Prior research  has established the basis for the development of the proposed protocol. The following section will focus on the most relevant literature and provide an overview of the current state of the art concerning stealth addresses.

Stealth addresses were first introduced to the blockchain domain by an anonymous entity dubbed ``bytecoin'' in April 2011. Subsequently, van Saberhagen and Todd put forward more refined stealth address protocols in 2013 and 2014, respectively~\cite{vansaber_2013, todd_2014}. These protocols laid the groundwork for the DKSAP implemented in the Monero blockchain upon its launch in 2014~\cite{getmonero}.
Since DKSAP's inception, numerous researchers have sought to extend the capabilities of the stealth addressing protocol and introduce additional features and functionality.

Courtois and Mercer~\cite{Courtois_2017} provide an overview of the development history of stealth addresses. Furthermore, the authors introduce multiple different spending keys to the DKSAP, improving its resistance to attacks such as the "bad random attack" or compromised keys. Their proposed protocol comes at the cost of requiring the users to manage multiple different spending keys.

Fan~\cite{Fan2018} improve the DKSAP, enabling sender and receiver pairs to use their generated Diffie Hellman secret multiple times with an increasing counter, enabling the protocol users faster parsing. Their approach is based on a similar idea as TLS and achieves performance gains of at least 50\% compared to the standard DKSAP.

Fan \emph{et al.}~\cite{Fan_2019} improve the DKSAP by reducing storage costs by reducing the number of keys required from two to one while using a bilinear mapping to maintain the desired properties of the DKSAP. The authors demonstrated notable efficiency gains by reducing the number of key pairs stored.

Liu \emph{et al.}~\cite{Liu_2019} implement stealth addresses together with ring signatures to define a confidential layer within a cryptocurrency system.
The authors use a lattice-based protocol to fully shield the information about the sender and recipient of a transaction.

Feng \emph{et al.}~\cite{feng2020} propose a stealth address protocol that does not require additional information to be published with every stealth address transaction, allowing such transactions to look like common transactions. The authors use the number of transactions between certain peers instead of generating a Diffie-Hellman secret for the stealth address generation process. This comes with the requirement for users to parse every transaction recorded on a blockchain.

Lee and Song~\cite{Lee_2021} use a stealth address protocol and ring signatures to implement confidential transactions on an Ethereum private network. The authors focus on the exchange of healthcare information and further analyze the security of their protocol using threat models.

Mohideen and Kumar~\cite{abdulkader2022privacy} build on top of the protocol of~\cite{feng2020} and use the transaction ID of the most recent P2P transaction between two entities instead of the number of transactions in the stealth address generation process. The authors argue that without the need to attach information to the stealth address, related protocols become more censorship-resistant and lightweight.

In summation, previous research underscores a substantial acceptance of the DKSAP. Several studies focused on reducing parsing time for recipients by introducing efficient strategies, such as deterministic rules that dictate the computation of stealth addresses between two parties based on an initially generated Diffie-Hellman secret or adopt sophisticated cryptographic algorithms such as bilinear mappings~\cite{Fan2018, Fan_2019}.

Furthermore, mitigating the problem of detectability in stealth address transactions can be achieved by refraining from publishing any extra information alongside stealth address transactions. However, this approach entails a significant drawback for blockchains that handle a large volume of transactions, in addition to stealth address transactions, as it requires parsing each transaction recorded.


\section{Background on Blockchain Privacy}\label{sec:background}

Privacy remains a primary concern within the realm of public blockchains. The inherent transparency of these systems may jeopardize users' privacy when conducting financial transactions or other sensitive interactions. To address this issue, blockchain developers have attempted to develop privacy-enhancing protocols that provide unlinkability and untraceability or focus exclusively on the former. In this context, we adhere to the definitions established in the CryptoNote whitepaper to define ``unlinkability'' and ``untraceability''. Per this reference, unlinkability is characterized as the inability to verify that two outgoing transactions are directed to the same recipient. Untraceability, on the other hand, is the inability to pinpoint the sender of a transaction from a group of potential senders.

\begin{description}[leftmargin=0cm,labelindent=\parindent, itemsep=0.5em]
\item[ZK-SNARKs. ]
There have been numerous efforts to bring confidential transactions to public ledgers such as Bitcoin and Ethereum, including the use of ZK-SNARKs ("Zero-Knowledge Succinct Non-Interactive Argument of Knowledge")~\cite{zcash, bowe2017scalable, Banerjee_2020, Guan_2022}. ZK-SNARKs enable a user to prove certain information without disclosing that information, which allows for the possibility of depositing funds into a Smart Contract using one pseudonym and then withdrawing those funds by proving the deposit under a different pseudonym without disclosing which deposit was referenced for the withdrawal. 
In addition to promoting enhanced scalability in the blockchain, this technology is implemented on Ethereum through privacy-enhancing tools like Tornado Cash or Privacy Pools and in Zero-Knowledge rollup platforms such as Aztec. This technology ensures both untraceability and unlinkability, thereby offering a robust means of preserving privacy.~\cite{pertsev2019tornado, ameen_soliman_2022, williamson2018aztec}.

\item[CoinJoin. ]
Chaumian CoinJoin is a privacy-enhancing technology used on UTXO-based blockchains that ensures the untraceability and unlinkability of transactions. CoinJoin is a process in which multiple users combine their UTXOs into a single, larger transaction. This consolidation complicates the task of an external observer trying to correlate input addresses (the senders) with output addresses (the recipients)~\cite{maxwell_2013, Ficsor_2017}. CoinJoins have been implemented in applications such as Wasabi Wallet, Samurai Wallet, and JoinMarket~\cite{deuber2021coinjoin}.
Blind signatures are employed to guarantee that the central coordinator cannot link the input and output addresses of the participants. Applied correctly, CoinJoins prevent the central coordinator or any other third party from tracing the flow of funds and de-anonymize users~\cite{chaum1983}.

\item[Stealth Addresses. ]
Stealth addresses are a privacy-enhancing solution that hides the recipient of a transaction and therefore prevents third parties from linking the interacting parties. By enabling senders to create a new stealth address for the recipient in a non-interactive manner, such protocols can provide unlinkability. While stealth addresses can be implemented on the application layer of programmable blockchains, some projects, such as Monero~\cite{getmonero}, opted to integrate them into the core protocol. Furthermore, stealth addresses can be employed on UTXO and account-based blockchain protocols.

One key difference between stealth addresses and ZK-SNARKs is the extent of privacy that can be achieved. ZK-SNARK-based privacy applications are used to prove information (cf. ``ownership'') of an asset without necessarily possessing that asset at the time. This allows for commingling funds with those of other users and consequently eliminating discernible on-chain traces. Stealth addresses obfuscate the recipient's ownership within a transaction by employing newly generated pseudonymous addresses. The funds remain traceable as funds are not commingled with third-party assets. This distinction is essential to consider when evaluating the suitability of these privacy-enhancing solutions for different use cases.

Another distinction is the computational overhead: by the time of writing, ZK-SNARKs typically have a more significant overhead regarding computational resources and setup, while stealth addresses can be implemented with minimal impact using existing tools that modern blockchain platforms already provide.

Unlike CoinJoins or mixing pools, stealth addresses do not aim to obfuscate the on-chain visible flow of funds but instead hide the interaction between an identified sender and recipient. Consequently, external observers can trace the flow of funds to specific stealth addresses. However, observers cannot identify the individual or entity behind those recipient addresses. In contrast, CoinJoins provide an additional privacy layer by allowing users to conceal their identities within an anonymous group of CoinJoin participants. Therefore, CoinJoins offer more comprehensive anonymity than stealth addresses but rely on more user interaction and coordination.

Despite the mentioned variations, we assert that employing stealth addresses through elliptic curve mathematics offers a more lightweight and interoperable approach, making it accessible to a larger audience. Additionally, the inherent decentralization of stealth addresses contributes to the robustness of the protocol and further promotes the principles of autonomy and user privacy. This can potentially enhance the adoption of privacy features on public blockchains. Numerous applications, such as donations or payroll transactions, may not demand the high anonymity offered by notable ZK-SNARK-based tools or CoinJoin wallets. In particular, in situations that require Know Your Customer (KYC) procedures, stealth addresses present a more suitable solution.
Moreover, the lack of commingling means that users do not inadvertently help malicious parties anonymize ill-gotten assets by contributing to an extended anonymity set. This property makes stealth addresses particularly suitable for interactions where it is desired to refrain from helping malicious parties.

\end{description}

\section{Definition}\label{sec:definition}
The following sections outline the various components of our proposed protocol, BaseSAP. Given that our first implementation is based on elliptic curve (EC) cryptography, we introduce the foundational principles of elliptic curves. We then define our stealth address protocol, divided into \emph{address generation} and \emph{parsing} sections.

\subsection{Elliptic Curve Cryptography}
We define an elliptic curve $E$ over a finite field $F_{p}$ where $p$ is a 256-bit prime and present it in Weierstrass form as 
\begin{align*} 
y^2 = x^3 + ax + b \mid x, y \in F_{p}
\end{align*}
with $a$ and $b$ representing constants that determine the shape and position of the respective curve. The coordinates $(x, y)$ are points on the elliptic curve that can take any value within $F_{p}$ and form an Abelian group. This group structure allows us, given two points, e.g., $P$ and $Q$ to solve for $R$ by performing a binary operation called point addition, such that $R = P + Q$. For any points $P$ and $Q$ on the curve, we know $P + Q$ must also be on the curve. 

We denote lower and uppercase letters to scalars and points on the curve, respectively. The scalars $p$ and $q$ are random integers of size $n$, such that $p$, $q\in \{0,1\}^{n}$. It is common for private keys, such as those used in the Secp256k1 curve, to have a size of 256 bits~\cite{secp256k1}. 
An EC multiplication of the point $P$ by the scalar $n$ can be done by repeatedly performing additions of the point along the curve, such that $n \times P= P_{i} + \dots P_{i+n}$. Another property of the point addition operation on an EC group is that it is commutative, meaning for all points, e.g., $P$ and $Q$, 
$Q + P = P + Q$. The EC has a generator point $G$, representing a fixed curve point. A public key can be derived by multiplying a scalar with the generator point, $P = p \times G$.

We denote the ``point at infinity'' $\mathcal {O}$ as the identity element of the EC arithmetic, such that \mbox{$\mathcal{O} + \mathcal{O} = \mathcal{O}$} and \mbox{$P + \mathcal{O} = P$}.
Finally, for every point $P$ on the elliptic curve, there exists an inverse point such that $(-P) + P = \mathcal{O}$.

The Standards for Efficient Cryptography (SEC) is a set of standardized elliptic curves proposed for use in cryptography. These curves are designated as ``SEC curves'' and are intended to provide a standard set of curves for use in various cryptographic applications.

One of the most well-known SEC curves is Secp256k1, which is defined by the equation $y^2 = x^3 + 7 \pmod{p}$, where $p = 2^{256} - 2^{32} - 977$. This curve has a prime order $n$ of approximately $2^{256}$, and it is used as the basis for the Bitcoin Elliptic Curve Digital Signature Algorithm (ECDSA)~\cite{secp256k1}.

Secp256k1 has several attractive properties, including a large prime order and efficient arithmetic, making it well-suited for use in cryptocurrencies. It has been widely adopted in various applications, including blockchain technologies, IoT, and secure communication protocols~\cite{Takahashi_2019, zhai2019research, Mehrabi_2020}.

Our proposed stealth addresses protocol is designed to be agnostic to the specific elliptic curve employed, although the initial implementation utilizes the Secp256k1 curve from the SEC set.

\subsection{Stealth Address Protocol}
The following section is divided into different parts. First, we explicate how users generate a stealth address using the Improved Stealth Address Protocol (ISAP), described by Todd~\cite{todd_2014}, when they want to perform a transaction. Second, we introduce the DKSAP, described by van Saberhagen~\cite{vansaber_2013}, primarily focusing on the parsing process which the receiver or third-party providers can perform.

\begin{description}[leftmargin=0cm,labelindent=\parindent, itemsep=0.5em]
\item[Stealth Address Generation. ]
For the stealth address generation, we define two independent parties, the sender $\mathcal{C}$ (cf. \emph{caller}) and the recipient $R$, who both have access to a cryptographic key pair, $(p, P)$ and $(r, R)$.
We assume the public key of the recipient $R$ is published and known to the respective sender. Furthermore, it is important to note that the sender uses an ephemeral key pair, $(p, P)$, which is randomly generated for each transaction using the stealth address protocol instead of using a key pair with a public key directly linked to their identity.

The sender generates a shared secret using the Elliptic Curve Diffie-Hellman (ECDH) protocol to derive a stealth address for interaction with the recipient. A stealth address is generated by adding the point obtained from multiplying the Diffie-Hellman (DH) secret with the generator point to the recipient's public key. The sender performs the following steps for this process:
\begin{enumerate}
 \item Generate an ephemeral key pair $(p, P)$ and publish the coordinates $P$.
  \item Multiply the randomly generated ephemeral private key with the recipient's public key: $k = p \times R$. This creates the DH secret, such that  $k = r \times P = p \times R = r \times p \times G$.
  \item Hash the shared secret $k_{h} = h(k)$, with $h$ representing a cryptographic hash function $h: \mathcal{X} \rightarrow \mathcal{Y}$.
  \item Multiply the hashed shared secret with the generator point $\mathcal{K}_{h} = k_{h} \times G$.
  \item Add the result of (4) to the recipient public key: \mbox{$R_{st} = \mathcal{K}_{h} + R$}.
\end{enumerate}
Let $R_{st} \in E(F)$ denote the point which party $\mathcal{C}$ uses as a stealth address for $R$. It is important to note that there is no direct link between the two parties, and to external observers, it appears as if $\mathcal{C}$ is interacting with a random account unrelated to $R$.
Furthermore, it is important to recognize that the sender can be confident that only $R$, as the owner of $r$, can access $R_{st}$ by deriving the private key $r_{st}$.

\item[Stealth Address Parsing. ]
The stealth address parsing process allows potential transaction recipients to locate their stealth address and obtain the private key required to access it. This procedure requires that every potential recipient conduct parsing across the complete set of published ephemeral public keys, denoted by $\mathcal{A} = \{P_i, \dots, P_n\}$. The total number of unique stealth addresses generated via the protocol is represented by $|\mathcal{A}|$. To conduct parsing, a potential recipient must first gather the set of all existing ephemeral public keys and then perform the subsequent steps on each $P \in \mathcal{A}$:

\begin{enumerate}
  \item Multiply $P$ with the private key $r$: $k = r \times P$.
  \item Hash the derived shared secret $k_{h} = h(k)$.
  \item Add the result of (2) to the own private key: \mbox{$r_{st} = k_{h} + r$}.
  \item Multiply the result of (3) with the generator point to derive the stealth public key: \mbox{$R_{st} = r_{st} \times G$}.
  \item Hash the stealth public key and take the least significant 20 bytes to derive the address: $R^{addr}_{st} = h(R_{st})\textbf{[}$-$20:\textbf{]}$.
\end{enumerate}
Upon deriving the point $R_{st}$, the recipient can determine whether $R^{addr}_{st}$ has been the recipient of the transaction or whether $R^{addr}_{st}$ received any assets. If the check is successful, the receiver may store the private key $r_{st}$.

To conclude, the protocol leverages the fact that $k_{h} \times G + P =  (k_{h}+p) \times G$. This allows for deriving a stealth address through two different paths, while only the recipient can generate the private key for the stealth address.

\item[Dual-key scheme. ]
Stealth addresses on Ethereum require the recipient to use their private key $p$ during the process. This has important implications for both security and user experience. First, users may encounter situations where they need to use their private keys for operations outside of their cold storage, which poses significant security risks. Second, users cannot delegate the parsing process to a third-party service as it involves sharing the private key and compromising its confidentiality. Therefore, users must perform the parsing process themselves in a local environment.

\begin{figure}[t]
\includegraphics[width=3.5in]{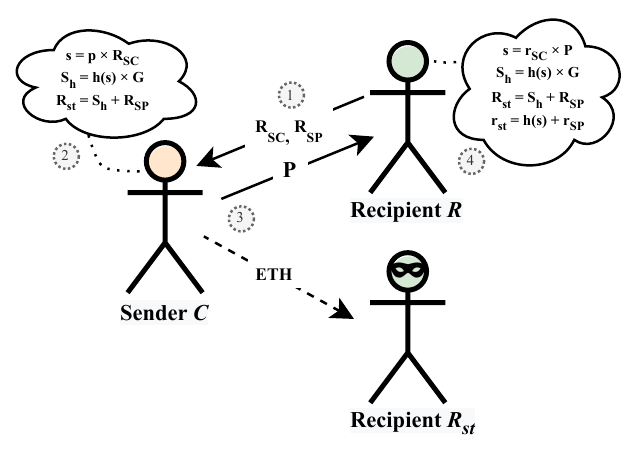}
\centering
\caption{ISAP + DKSAP: (1) sender gets public keys of the recipient, (2) generates the stealth address, and (3) sends to stealth address and publishes an announcement; (4) recipient uses the announcement to derive the private key that unlocks the stealth address.}
\label{Fig:basicprotocol}
\end{figure}

Researchers have developed a solution to these issues, the DKSAP, which improves both the user experience and security~\cite{feng2020, Fan2018, Courtois_2017}.
The DKSAP is an extension of the ISAP and introduces an additional key pair exclusively used for the parsing process. Recipients have two key pairs \mbox{--- scanning and spending keys ---} represented as $(r_{\mathcal{SC}}, R_{\mathcal{SC}})$ and $(r_{\mathcal{SP}}, R_{\mathcal{SP}})$, respectively. Equipping the recipient with two separate key pairs, the scanning key pair, which is still used in the DH secret generation, can be partially separated from the stealth address generation. To use the DKSAP, the sender needs to follow these steps:

\begin{enumerate}
  \item Multiply the randomly generated ephemeral private key with the scanning public key of the recipient: $k = p \times R_{\mathcal{SC}}$.
  \item Hash the shared secret $k_{h} = h(k)$ and multiply the result with the generator point $\mathcal{K}_{h} = k_{h} \times G$.
  \item Add the result of (2) to the recipient spending public key: \mbox{$R_{st} = \mathcal{K}_{h} + R_{\mathcal{SP}}$}.
\end{enumerate}
Henceforth, the recipient has two options for finding the respective stealth address $R_{st}$. First, the recipient can calculate the DH secret by multiplying the scanning private key $r_{\mathcal{SC}}$ with the ephemeral public key $P \in \mathcal{A}$. Having the DH secret, the recipient can derive the stealth address by hashing it, multiplying the hash with the generator point, and adding the result to the spending public key. Second, the recipient can add the DH secret to the spending private key and multiply the result with the generator point for deriving the stealth address $R_{st}$:
\begin{align*} 
R_{\mathcal{SP}} + h(r_{\mathcal{SC}} \times P) \times G = (r_{\mathcal{SP}} + h(r_{\mathcal{SC}} \times P)) \times G.
\end{align*} 
It is important to mention that the recipient can share the scanning private key $r_{\mathcal{SC}}$ with a third-party parsing provider without compromising the spending private key. Using the scanning key, the parsing provider can take on the parsing task and notify users when an incoming stealth address transaction occurs. However, without access to the spending private key $r_{\mathcal{SP}}$, parsing providers cannot access the stealth address.
\end{description}

\section{BaseSAP Protocol}\label{sec:stealthaddress}

In the following, we describe our proposed stealth address protocol in detail. This involves on-chain key management solutions, stealth address transaction routing, and specific efficiency improvements we propose to the DKSAP. 

Our protocol is designed to operate on the application layer of programmable blockchains such as Ethereum and does not require integration with the core protocol layer of a blockchain. After deployment, BaseSAP operates autonomously, eliminating the possibility of user interference or censorship of any party. 

The proposed protocol can serve as a foundation for various implementations to build upon it and leverage the modular basis.
BaseSAP comprises a single singleton contract, the \emph{Announcer} contract, which enables users to publish the ephemeral public keys at a central place.

For the initial implementation of an improved version of a DKSAP, built on top of BaseSAP, we propose a \emph{Registry} contract that serves as a central repository for stealth meta-addresses associated with registered users. 

We practically implement our proposed protocol on both the Ethereum Goerli and Sepolia testnets to analyze the performance of the proposed protocol under practical conditions and enable the community to engage with it.

\subsection{Announcer contract}\label{sec:transactions}
To interact with a user's stealth address, the sender must first obtain the recipient's public key and then generate a stealth address. This process involves the sender using their own randomly generated ephemeral private key and the recipient's public key to derive the stealth address.

To enable recipients to detect their stealth addresses, senders must publicly announce their ephemeral public keys. Adversaries cannot exploit this announcement to compromise a recipient's privacy, as they cannot recreate the necessary Diffie-Hellman (DH) secret for generating the stealth address.

\begin{figure}[t]
\includegraphics[width=2.5in]{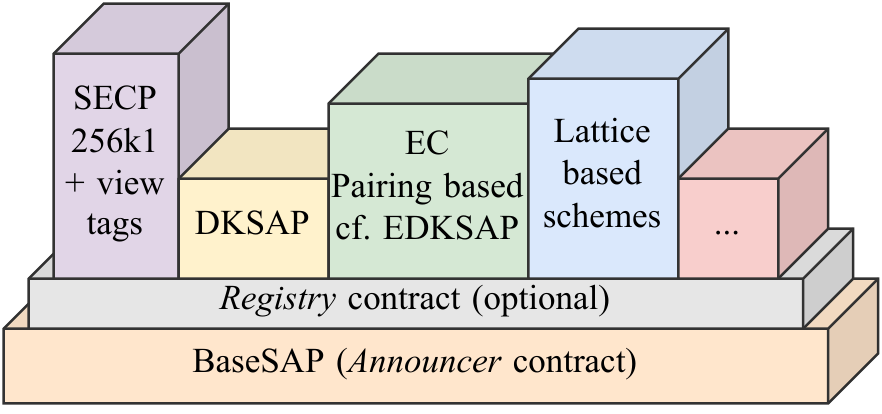}
\centering
\caption{Modular property: BaseSAP enables different stealth address schemes to build on top of it and leverage the modularity, interoperability and trust of the underlying foundational protocol.}
\label{Fig:modsap}
\end{figure}

The \emph{Announcer} contract emits announcements from a central location to which users can subscribe. Unlike~\cite{feng2020}, where the objective was to have stealth address transactions mimic regular transactions, providing additional information --- particularly the ephemeral public key --- in conjunction with each stealth address transaction is necessary. Without attaching additional information to a stealth address interaction, the protocol would require users to parse the entire ledger~\cite{feng2020}. However, by emitting events containing that information, users can utilize the existing Bloom filters on the Ethereum blockchain to identify the set of transactions related to stealth address interactions.

The \emph{Announcer} contract is designed to be agnostic to the cryptographic scheme used, enabling different implementations of various cryptographic schemes to share the same source of event emissions. This means that the same \emph{Announcer} contract can be used for multiple distinct implementations without needing modification or adjustment. This essential characteristic is illustrated in Figure~\ref{Fig:modsap}.

\begin{description}[leftmargin=0cm,labelindent=\parindent, itemsep=0.5em]
\item[Announcements. ]
Employing BaseSAP, the transaction used to transfer assets also serves to broadcast the announcement, containing the ephemeral public key, to the public, making the information accessible to the recipient. To facilitate this feature, a lightweight \emph{Announcer} contract is used (refer to Listing~\ref{lst:announcement}). The \emph{Announcer} contract can be called by anyone to emit additional information along with a transaction.

External Owned Accounts (EOAs) can call the \emph{Announcer} contract using regular transactions, while contracts can interact with the \emph{Announcer} through internal calls to execute the \emph{announce} function. As shown in Listing~\ref{lst:announcement}, the caller can provide five parameters to the function:
\begin{enumerate}
 \item scheme ID --- to specify the cryptographic scheme that was used. In BaseSAP, the Secp256k1 implementation is assigned the identifier number 1. An incrementing number is assigned for subsequent implementations.
  \item stealth address --- the address of the transaction recipient.
  \item caller address --- the address of the calling entity, the sender.
  \item ephemeral public key --- compressed public key derived from the randomly generated ephemeral private key.
  \item metadata --- arbitrary information that may be helpful to the recipient in identifying the particular interaction between the sender and the stealth address. For \mbox{ERC-20} tokens, the metadata field may include the four most significant bytes of the method id, 20 byte for the (token) contract address, and 32 bytes carrying the amount transferred. For ERC-721/ERC-1155 contracts, the metadata field may include the token ID instead of the amount transferred. For regular ETH transfers, the metadata field may remain empty.
\end{enumerate}
Incorporating an extra metadata field within the announcement enables recipients to verify receipt of an asset, including the amount transferred and  the specific token involved. Moreover, by incorporating the method ID in the metadata, recipients can pinpoint the contract interaction involving their stealth addresses. As a result, well-known interactions, such as token approvals or mints that carry the right to execute on a certain state, are compatible with stealth addresses. 
Consequently, the recipient isn't obligated to perform extra Remote Procedure Calls (RPCs) to obtain information about the nature or quantity of an asset or right received.

Considering the metadata field's dynamic size, it can also be employed to incorporate additional features for enhancing parsing at a later stage or to include more information that might be needed for future token standards.

\begin{listing}[t!]
\begin{minted}[xleftmargin=9pt,tabsize=1,linenos=true,breaklines=true,breakanywhere=true, fontsize=\scriptsize,numbersep=4pt,breaksymbol=,breakindent=30pt ]{solidity}
pragma solidity ^0.8.0;

/// @notice Announcer emitting the Announcement event.
interface BaseSAPAnnouncer {

  /// @notice Emitted when interacting with a stealth address.
  event Announcement (
    uint256 indexed schemeId, 
    address indexed stealthAddress, 
    address indexed caller, 
    bytes ephemeralPubKey, 
    bytes metadata
  );
  
  /// @notice To be called when interacting with a stealth address.
  function announce (
    uint256 schemeId, 
    address stealthAddress, 
    bytes memory ephemeralPubKey, 
    bytes memory metadata
  )
    external
  {
    emit Announcement(
      schemeId, 
      stealthAddress, 
      msg.sender,
      ephemeralPubKey, 
      metadata
    );
  }
}
\end{minted}
\caption{Announcer Contract Interface}
\label{lst:announcement}
\end{listing}

\item[Costs. ]
Executing the \emph{announce} function with the parameters used for an ERC-20 transfer consumes approximately $35$,$492$ units of gas on the Ethereum blockchain. If the metadata field is left empty, as it would be in the case of Ether transfers, the gas usage is reduced to $34$,$057$ units of gas. Assuming a gas price of $10$ $gwei$ and a price for $1$ Ether (ETH) of $2$,$000$ US dollars, with $1$ ETH = $ 10e9$ gwei = $ 10e18$ wei, the cost of calling the \emph{announce} function for Ether transfers is approximately $0.68$ USD, or $0.00034057$ ETH. On layer-2 rollup platforms such as Optimism\footnote{https://www.optimism.io/} or Arbitrum\footnote{https://arbitrum.io/}, the costs associated with the announcement are effectively negligible.

Compressing the ephemeral public key to 33 bytes can reduce the gas consumption to $35$,$064$ units. Without the metadata field, the emission of the announcement consumes $33$,$629$ units of gas. From these figures, we can deduce that applying public key compression results in cost savings of $1.21\%$ for non-empty metadata emissions and $1.26\%$ for ETH transactions that do not require metadata.

\end{description}

\subsection{Stealt Meta-Address Format}\label{sec:stealth-meta-address}
In the design of the DKSAP, the recipient has two separate key pairs, the spending $\mathcal{SP}$ and the scanning keys $\mathcal{SC}$. We combine the two public keys to generate the \emph{stealth meta-address}, enabling a more intuitive way  for users to interact with each other. In Secp256k1, the public keys $PKs$ can be compressed to 33 bytes each, denoted as $PK_{comp}$. Consequently, our proposed protocol uses the following format for the stealth meta-address:
\begin{align*} 
st:\langle\emph{chainId}\rangle~:~0x\langle PK_{comp}^\mathcal{SP}\rangle \langle PK_{comp}^\mathcal{SC} \rangle
\end{align*}

The ``st'' prefix indicates that the following address refers to a stealth meta-address. The ``chainId'' parameter distinguishes blockchain-specific addresses that the corresponding recipient is open to interact with via stealth addresses. Within the Ethereum ecosystem, chain IDs have been formalized in the ERC-3770.

To compress the public key, we store only the x-coordinate of the public key point and prefix it with either 0x2 or 0x3, depending on whether the y-coordinate is positive or negative, respectively. The stealth meta-address can be shared through off-chain communication channels or made publicly available on the blockchain. This way, any individual can generate stealth addresses on behalf of the user, thereby facilitating their interaction.

For cryptographic schemes where compressing the public key is not feasible, the full public keys can be used instead.

\subsection{Secp256k1 Implementation}\label{sec:parsing}

In the following, we propose efficiency improvements to the DKSAP to make it more viable for implementation on blockchain platforms like Ethereum. Our analysis of the existing protocols highlights two deficiencies that impede their practical usage. First, we observe that the parsing process required for every potential recipient to decode every announcement can be excessively time-consuming. Second, we note that the announcement, which merely contains the ephemeral public key, does not offer sufficient information for recipients to identify the relevant assets and rights in a stealth interaction.

To remedy these limitations, we focus on enhancing the efficiency and flexibility of the recipients by modifying the announcement and publication process and introducing a ``view tag'' approach. We intend to enhance the overall functionality of the protocol, thereby facilitating its application in blockchain environments.

\begin{description}[leftmargin=0cm,labelindent=\parindent, itemsep=0.5em]
\item[Announcement. ]
The set of announcements, denoted by $\mathcal{A} \mid a \in \mathcal{A}$, contains information that enables prospective recipients to identify themselves as the intended recipient of a stealth address transaction and locate the corresponding stealth address. Generally, the recipient utilizes the ephemeral public key disclosed by the sender to compute the stealth address and validate that it corresponds with the recipient address of the transaction. This procedure allows users to confirm that they are indeed the rightful recipients of the transaction. This process involves two RPC requests --- one for procuring the announcement and another for retrieving the transaction recipient.

To refine the parsing process and enhance its flexibility, we propose integrating the stealth address $R^{addr}_{st}$ into the announcement $a$ that is emitted for every stealth address transaction. This enables a direct comparison between the derived stealth address $R^{addr}_{st}$ and the address listed in the announcement $R^{x}_{st}$', enabling the recipient to determine if they are the intended recipient without the need to initiate supplementary RPC requests to obtain transaction recipients or query balances for different assets on the derived stealth address. The condition \mbox{$R^{x}_{st} == R^{x}_{st}$'} confirms whether the parsing user is the intended recipient. By consolidating all the vital information the recipient needs within the announcement, we obviate the need for the recipient to initiate an extra RPC call.

\item[View Tag. ]
View tags represent a technique employed within the Monero blockchain protocol, allowing recipients of stealth address transactions to bypass certain steps in the parsing process under specific conditions~\cite{ukoe, getmonero}. Rather than computing the stealth address and comparing it to the address in the announcement, the recipient can hash the DH secret and compare the most significant $n$ bytes with the ``view tag'' documented in the announcement. In this case, $n$ can be kept very small, so setting $n=1$ means that a full derivation of the stealth address must only be attempted 1/256 of the time at the cost of only a single-byte view tag. To construct the view tag, senders use their ephemeral key pair $(p, P)$ to compute the hashed DH secret $k_h = h(p \times R_\mathcal{SC})$ and select the most significant $n$ bytes of $k_h$. The resultant view tag $\mathcal{Q}$, where $\mathcal{Q} = k_h\textbf{[}$:$n\textbf{]}$, is disclosed alongside the stealth address transaction.

The recipient, possessing the scanning key pair $(r_\mathcal{SC}, R_\mathcal{SC})$, can also compute the view tag by following the same procedure, $\mathcal{Q}' = h(r_\mathcal{SC} \times P)\textbf{[}$:$n\textbf{]}$, and compare it to the view tag listed in the announcement $\mathcal{Q} == \mathcal{Q}'$. If the view tags do not match, the parsing user can omit every subsequent operation for the current announcement and advance to the next one. 
It's worth noting that exposing $n$ bytes of the hashed DH secret affects the users' privacy, as attackers may attempt to brute-force a user's stealth address by applying the view tag to potential recipients. Nevertheless, such attacks are likely to be successful only for sufficiently large $n$.

The parsing process then involves the following steps on each announcement $(P, R^{addr}_{st}, \mathcal{Q}) \in a \mid \forall a \in \mathcal{A} $:

\begin{enumerate}
  \item Multiply $P$ with the scanning private key $r_{\mathcal{SC}}$: \mbox{$k = r_{\mathcal{SC}} \times P$}.
  \item Hash the derived shared secret $k_{h} = h(k)$. 
  \item Derive the view tag $\mathcal{Q}$'$ = k_{h}\textbf{[}$:$n\textbf{]}$
  \item Compare the derived view tag with the one in the announcement $\mathcal{Q} == \mathcal{Q}'$. 
  \item Only if the view tag matches, the recipient continues to compute the stealth address and compare it to the address logged $R^{addr}_{st}$'$ = h((k_{h} + r_{\mathcal{SP}}) \times G)\textbf{[}$-$20:\textbf{]} = R^{addr}_{st}$.
 
\end{enumerate}

Similar to Monero we use view tags that are 1 byte in size and prepend them to the metadata field, taking up the first byte of it.

\begin{figure}[t]
\includegraphics[width=3.5in]{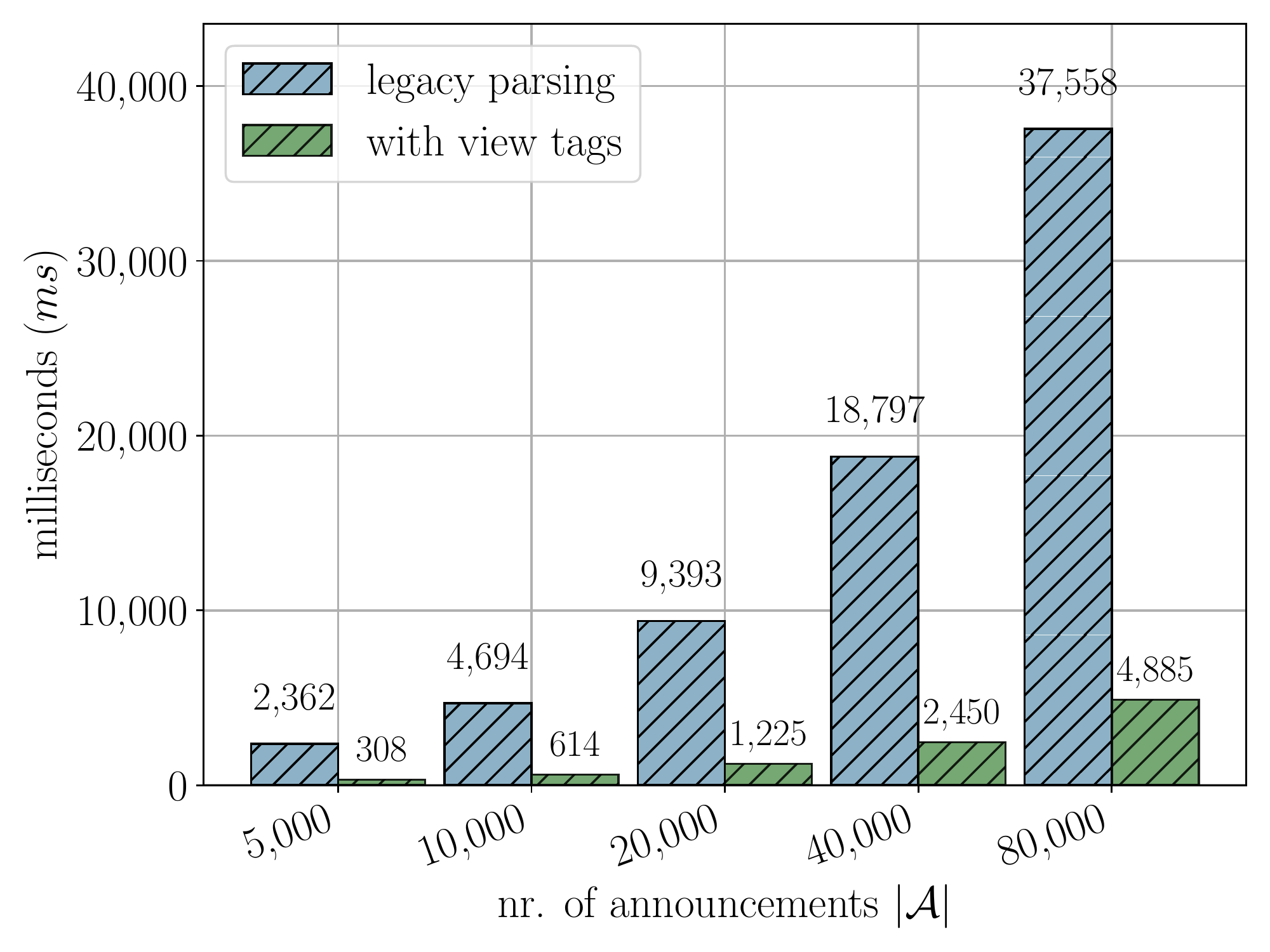}
\centering
\caption{View tag improvements: Conventional parsing (without employing view tags) versus an upgraded variant using view tags. In summary, the view tags approach operates approx. $7.6$ times more efficient concerning parsing time than conventional parsing. Both algorithms have a complexity of $O(n)$.}
\label{Fig:viewtagalgo}
\end{figure}

Figure~\ref{Fig:viewtagalgo} presents simulations that juxtapose the efficiency gains achieved by implementing view tags, as measured by parsing time. We compare the view tag approach to the conventional DKSAP method (cf. ``legacy parsing''). The experiments were carried out on a machine with a 10-core CPU Apple M1 Max chip using Node.js/JavaScript, with the \emph{elliptic.js}\footnote{https://github.com/indutny/elliptic} and \emph{js-sha3.js}\footnote{https://github.com/emn178/js-sha3} libraries employed for EC operations and hash functions, respectively. No efficiency improvements through multiprocessing were leveraged.
Legacy parsing requires the recipients of a stealth address transaction to perform the following operations to ascertain if they were the recipients of stealth address transactions:

\begin{itemize}

\item 2x ecMUL --- DH secret \& hashed secret times generator,
\item 2x HASH --- hash of DH secret \& address derivation,
\item 1x ecADD --- deriving the stealth address

\end{itemize}

Adopting \emph{view tags} significantly decreases parsing time by approximately $86.84$\%. In most cases, users are only required to perform a single EC multiplication operation (ecMUL) and a single hash operation (HASH), thus eliminating the necessity for an additional ecMUL, ecADD, and HASH. With a 1-byte view tag, the likelihood that users can bypass the remaining computations after hashing the shared secret is 
$1/256$
. This suggests that users can almost certainly bypass the aforementioned three operations for most announcements. The realized reduction in parsing time comes with a significant positive impact on the user experience. As displayed in Figure~\ref{Fig:viewtagalgo}, for $80$,$000$ announcements, \emph{view tags} enable the reduction of the parsing time from $37.56$ to $4.89$ seconds.

\end{description}

\subsection{Public Key Management}\label{sec:keymanagement}
Considering the usage of dual-key mechanisms, we advocate for integrating a key management solution that facilitates blockchain users to store their stealth meta-addresses in a predefined location publicly. Absent a dual-key configuration, a central repository for storing public keys wouldn't be necessary. Users could alternatively derive another user's public key by extracting it from a transaction that the latter has signed. It's crucial to mention that the stealth address protocol could still be employed even without a central repository. Thus, any key management solution can be built atop the fundamental protocol.

We design a fully autonomous and lightweight registry contract to maintain a record of registered users and their corresponding stealth meta-addresses. This contract predominantly consists of getter and setter methods that assist users in registering their stealth meta-addresses on the blockchain or retrieving those of others. Moreover, the registry permits users to register a stealth meta-address on behalf of another user by providing a valid signature from the respective registrant. Finally, an event is broadcasted each time a user registers a new stealth meta-address.

\begin{listing}
\begin{minted}[xleftmargin=9pt,tabsize=1,linenos=true,breaklines=true,breakanywhere=true, fontsize=\scriptsize,numbersep=4pt,breaksymbol=,breakindent=30pt ]{solidity}
pragma solidity ^0.8.0;

interface IERC5564Registry {

  /// @dev Emitted when a registrant updates their stealth meta-address.
  event StealthMetaAddressSet(
    bytes indexed registrant, 
    uint256 indexed scheme, 
    bytes stealthMetaAddress
  );

  /// @notice Maps a registrant's identifier to the scheme to the stealth meta-address. 
  mapping(bytes => mapping(uint256 => bytes)) public;

  /// @notice Sets the caller's stealth meta-address for the given stealth address scheme.
  function registerKeys(
    uint256 scheme, 
    bytes memory stealthMetaAddress
  ) external;
  
  /// @notice Sets the `registrant`s stealth meta-address for the given scheme.
  function registerKeysOnBehalf(
    address registrant,
    uint256 scheme,
    bytes memory signature,
    bytes memory stealthMetaAddress
  ) external;
}
\end{minted}
\caption{Registry Contract Interface}
\label{lst:registry}
\end{listing}

Our proposed registry contract includes dynamic size storage slots for the stealth meta-address to ensure compatibility with various elliptic curves and cryptographic schemes. This allows for the construction of supplementary stealth address implementations atop the existing framework, capitalizing on the benefits of a shared registry. Users can register distinct stealth meta-addresses for different cryptographic schemes by specifying a scheme ID. For instance, a user could register one stealth meta-address with an elliptic curve $E(F)$ and another for the curve \mbox{$E(F')$ $\mid F \neq F'$}, thus avoiding conflicts. This provision ensures compatibility with future cryptographic methods.

\newcommand{\pluseq}{\mathrel{+}=}
\newcommand{\subeq}{\mathrel{-}=}
\begin{algorithm}
\setstretch{1.2}
\caption{--- Register Stealth Meta-Address}
\label{alg:keysregister}
\begin{algorithmic}[1]
 \renewcommand{\algorithmicrequire}{\textbf{Input}}
\REQUIRE \textbf{I:} Scheme ID $id$\\
\REQUIRE \textbf{II} Stealth Meta-Address $SMA$ in byte-format\\
\REQUIRE \textbf{III:} Caller/Signer $\mathcal{C}$\\
\REQUIRE \textbf{IV: (optional):}  Signature $sig$\\
 \vspace{0.15cm}
 \STATE $R^{x}_{\mathcal{SC}}, R^{x}_{\mathcal{SP}} \gets parse\_compressed\_pubkeys(SMA)$ 
  \STATE $R^{pre}_{\mathcal{SC}}, R^{pre}_{\mathcal{SP}} \gets R^{x}_{\mathcal{SC}}\textbf{[}0\textbf{]}, R^{x}_{\mathcal{SP}}\textbf{[}0\textbf{]}$ 
  \STATE $R^{x}_{\mathcal{SC}}, R^{x}_{\mathcal{SP}} \gets R^{x}_{\mathcal{SC}}\textbf{[}1$:$\textbf{]}, R^{x}_{\mathcal{SP}}\textbf{[}1$:$\textbf{]}$ 
 \STATE \textbf{assert} $R^{pre}_{\mathcal{SP}} == (2|3)$ \& $R^{pre}_{\mathcal{SC}} == (2|3)$
  \STATE \textbf{assert} $on\_curve(R^{x}_{\mathcal{SP}}, id)$ \& $on\_curve(R^{x}_{\mathcal{SC}}, id)$
  \IF{sig}
 \STATE registerKeysOnBehalf($id, \mathcal{C}, SMA, sig$)
 \ELSE
 \STATE registerKeys($id, \mathcal{C}, SMA$)
 \ENDIF
 \vspace{0.15cm}
\end{algorithmic}
\end{algorithm}

The \emph{registerKeys} function accepts the scheme ID, the stealth meta-address, and, optionally, a signature. The stealth meta-address includes a variable-sized field, allowing for managing diverse public key formats and sizes. As illustrated in Algorithm~\ref{alg:keysregister}, the registration process involves parsing the stealth meta-address to confirm that the two public keys are on the specified elliptic curve. This validation can be performed off-chain, hence circumventing unnecessary computations. For other cryptographic methods, other validation methods must be used.

Our proposed \emph{registerKeys} function registers the public keys through a mapping that associates the scheme ID with the stealth meta-address. This mapping is subsequently linked to the registrant's address in another mapping. Senders can interact with the registry to retrieve the stealth meta-address of another user by providing a scheme ID and the recipient's address.

\section{Security Implications}\label{sec:security}
In this section, we present an overview of the security implications that arise from our proposed protocol. Initially, we focus on DoS attack vectors that may compromise our protocol and put forth a cost function to guide the choice of appropriate DoS attack prevention measures. Our focus is on two different DoS attack prevention mechanisms: a \emph{toll} and a \emph{staking} system. After that, we direct our attention toward the privacy implications of the protocol, with a specific emphasis on the risks of user de-anonymization.

\subsection{Stealth Addresses and DoS attacks}\label{sec:dosattacks}
Ensuring the parsing process does not consume excessive time or CPU resources is vital. As previously mentioned, the parsing process involves several EC operations executed off-chain, circumventing blockchain-associated gas costs. However, this leaves the protocol vulnerable to DoS attacks, in which malicious actors flood the network with announcements. This forces users to carry out redundant EC operations on false announcements, leading to unnecessary consumption of computational resources. The costs of emitting an announcement may be lower than the costs of parsing it, which can result in inefficiencies and negatively impact the user experience by unnecessarily prolonging the parsing process.
The following section focuses on two different approaches for mitigating DoS attacks. We elaborate on both and highlight why a staking-based approach is better aligned with our purpose.

\begin{description}[leftmargin=0cm,labelindent=\parindent, itemsep=0.5em]
\item[Toll. ]

The DoS attack vector can be addressed by introducing a \emph{toll} system that accounts for the computational costs incurred during the parsing process. In particular, the \emph{toll} $\mathcal{T}$ may cover the parsing costs $c$, which are accrued until the hashed DH secret is derived. These costs comprise an EC multiplication $c_{mul} \in c$ and a hashing operation $c_{hash} \in c$. The \emph{toll} can be attached to the transaction and paid by the sender, who generates announcements and contributes to the parsing costs. This strategy ensures that the parsing costs associated with announcements are not exclusively shouldered by the recipients but also shared by the senders.

For the toll, we assume that:
\begin{align*} 
\mathcal{T} \geq c_{mul}+c_{hash},
\end{align*}
while the costs without using view tags can be described as $2(c_{mul}+c_{hash})+c_{add}$.
To ensures that adversaries pay at least for the effort imposed on a single individual stealth address recipient, the protocol must collect a fee of up to $\mathcal{T}$. 

The keccak-256 opcode, used with a 64 bytes input, costs 42 gas and is therefore negligible. Taking the EIP-196~\cite{reitwiessner2017eip196} precompiled contracts for the alt\_bn128 curve a reference for the cost of EC operations, we assume a gas usage of $40$,$000$ units for EC multiplications. Therefore, a toll of $40$,$000$ gas units might be suitable. Based on a gas price of 10 gwei, the toll would amount to 0.0004 ETH.

Indeed, the purpose of the \emph{toll} is to introduce a financial hurdle that makes spamming economically unviable rather than covering the entire costs of the parsing process. Therefore, it is possible to significantly reduce the \emph{toll} while still achieving the goal of discouraging DoS attacks. The specific value of the \emph{toll} should be carefully determined based on various factors such as network conditions, the overall cost structure of the protocol, and the desired level of protection against spamming. Finding the right balance enables establishing a cost-effective solution that effectively mitigates DoS attacks without imposing excessive financial burdens on legitimate users.

There are various ways to utilize the collected toll, but ensuring that it does not directly return to the originator to maintain the DoS attack protection is essential. One option for the proposed protocol is to send it to the coinbase address of the respective block, which is the address of the block proposer. This approach would distribute the \emph{toll} among block proposers, giving them an additional incentive to include stealth address transactions in blocks. Proposers then have an additional source of extractable value, encouraging them to prioritize stealth address transactions in block creation. However, the initiators of stealth address transactions may offset the expenses associated with the toll by reducing the gas price. Since block proposers are indifferent to whether the paid fee originates from transaction fees or payments to a block's coinbase, this strategy would effectively impede the toll's spam prevention.

Further research and analysis are needed to determine the optimal value of the \emph{toll} in different network environments and under varying circumstances. The same applies to the optimal use of the \emph{toll} to not introduce trust requirements or centralizing vectors through the back door.

\item[Staking. ]
It is worth mentioning that the dual-key setup of our proposed stealth address protocols enables users to share the private scanning key with third-party entities specializing in the parsing process. These parties can offer their services for a market price and come equipped with defense measures against DoS attacks. These measures can be based on specific heuristics that help identify spammers. As a result, third-party parsing providers can provide additional protection against DoS attacks targeting users, thereby ensuring the effectiveness and reliability of the parsing process.

A staking system can be implemented to equip parsing providers with an additional tool for managing spam and Sybil attacks. The \emph{Announcer} contract may permit users to stake an arbitrary amount of ETH and lock it within the contract. Parsing providers can then confirm if the sender of a stealth address transaction has staked some necessary collateral. If not, they can deprioritize the respective announcements of that sender when serving them to parsing users.

Analogous to the ERC-4337 standard, a \emph{MIN\_STAKE\_VALUE} and a \emph{MIN\_UNSTAKE\_DELAY} variable are established. The latter could be directly encoded into contracts communicating with the \emph{Announcer} contract and may be set to one day. The \emph{MIN\_STAKE\_VALUE} can be agreed upon off-chain and change over time. Theoretically, every parsing provider may independently set the minimum stake required for prioritization. 

We define the staking system as follows:

\begin{itemize}
  \item Let $\mathcal{A}$ be the set of all announcements, where 
  \begin{align*}
  A &= \{a_1, a_2, a_3, \dots, a_n\}.
  \end{align*}
  \vspace*{-1.5\baselineskip}%
  \item Let $U$ be the set of all users, where 
  \begin{align*}
  U &= \{u_1, u_2, u_3, \dots, u_m\}.
  \end{align*}
  \vspace*{-1.5\baselineskip}%
  \item For each user $u \in U$, let $D(u)$ be the amount of ETH deposited by user $u$.
  \item Let $F$ be the function that maps a prioritization factor $PF$ to users $F: u_i \rightarrow PF$.
\end{itemize}
We can define two priority factors, one based on the amount of ETH staked ($PF1$) and the other based on the number of announcements made by a user ($PF2$).

\begin{enumerate}
  \item \textbf{PF1: Staking priority factor}\\
  For each user $u_i \in U$ and their corresponding deposited ETH amount $D(u_i)$, we can define the staking priority factor ($PF1$) as:
  \begin{align*} 
PF1(u_i) = \emph{min}(D(u_i),\ \emph{MIN\_STAKE\_VALUE} )
\end{align*}

  Users staking more than the\emph{ MIN\_STAKE\_VALUE} are assigned a first prioritization factor equaling the \emph{MIN\_STAKE\_VALUE}

  \item \textbf{PF2: Announcement count priority factor}\\
  For each user $u_i \in U$ and the number of announcements made by $u_i$, the announcement count priority factor can be defined as:

  \[n(u_i) = |\{a_j \in \mathcal{A} : a_j \text{ is made by } u_i\}|\]

  We want to assign a higher priority to users who made fewer announcements to discourage spamming. We can define the announcement count priority factor ($PF2$) as:

  \[PF2(u_i) = \frac{1}{n(u_i)}\]

  Higher values of $PF2$ indicate higher priority for a user's announcements.
\end{enumerate}

Now, we can combine these two priority factors into a single prioritization factor ($PF$) for each user:

\[PF(u_i) = w_1 \cdot PF1(u_i) + w_2 \cdot PF2(u_i)\]

Where $w_1$ and $w_2$ are weights assigned to $PF1$ and $PF2$, respectively, to balance their importance in determining the overall priority. For example, if we want to prioritize ETH staked over the number of announcements made, we could assign $w_1 > w_2$. Initially, $w_1$ and $ w_2$ are set to 1.

Finally, parsing providers can order the list of announcements based on each user's computed $PF$ values, prioritizing announcements made by users with higher $PF$ values.

The proposed mechanism guarantees that the announcements from staking users are prioritized in the parsing process over those from users who did not stake. Furthermore, announcements from users with fewer announcements are given precedence.

It is worth noting that the staking-based DoS attack prevention is implemented on the parsing side, allowing parsing providers to manage spam more effectively. Spamming users can be disregarded or deprioritized when serving their announcements to parsing users. By requiring a stake for prioritization, Sybil attacks become inefficient, and the stake of known spammers can be traced. This prevents spammers from switching addresses to evade deprioritization.

In addition to preventing spam via Sybil attacks, the staking method does not impose any costs on the user side. The minimum required stake can be directly locked in the contract communicating with the \emph{Announcer} contract within the transaction that ultimately interacts with a stealth address. This allows for a seamless user experience.

Based on the reasons discussed, we deem the staking approach superior for our specific case instead of requiring a \emph{toll} for every stealth address transaction.
\end{description}

\subsection{Privacy Guarantees and De-anonymization}\label{sec:privacy-and-deanoymization}

Stealth addresses can provide users with an additional layer of confidentiality by allowing them to interact with certain entities without letting the public know. Several things must be considered to remain private as a stealth address recipient. Mistakes, such as compromising private keys, can cause de-anonymization or even the loss of user funds. 

\begin{description}[leftmargin=0cm,labelindent=\parindent, itemsep=0.5em]
\item[Commingled Funds. ]
In Ethereum de-anonymization studies, various heuristics are typically used to cluster addresses on a user basis~\cite{Chan2017, Chen2018, Lin_2020}. The use of stealth addresses on Ethereum has the potential to improve privacy by concealing the identity of the recipient of funds. However, commingling these funds, i.e., mixing them with other assets, can introduce risks to the overall privacy of the protocol. Particularly, when stealth address funds become intermingled with ``doxxed'' funds — assets already associated with a specific individual or entity through public records or other means — they forfeit the privacy benefits they initially offered. This intermingling can transpire during transactions involving withdrawals from a stealth address. Users who lack a comprehensive understanding of the privacy enhancements provided by stealth addresses may unintentionally transfer funds from a stealth address to an exposed, or ``doxxed,'' address. This practice can effectively erode the anonymity of the stealth recipient. Therefore, it is paramount for users to thoughtfully assess the risks associated with commingling funds destined for a stealth address.

\item[Transaction fee funding. ]
To cover the transaction fees associated with either spending an ERC-20 token or executing an approval right on the Ethereum blockchain, it becomes necessary for the recipient to furnish their stealth address with a minor quantity of ETH, avoiding the use of a publicly identifiable address. To circumvent this issue, the sender can append a small amount of ETH to each stealth address transaction, enabling the recipient to perform an on-chain action incurring a gas cost. By definition, this is not an issue for non-transferable assets, such as Soulbound tokens~\cite{weyl2022decentralized}. Alternatively, the recipient can fund the stealth address with anonymized ETH retrieved through another privacy tool, such as Tornado Cash, or via a trusted centralized exchange.

Another solution to the recipient's transaction fee funding problem involves entrusting specialized transaction aggregators, often known as ``searchers'' in the Miner Extractable Value (MEV) context. These intermediaries can provide users with the option of a one-time payment in exchange for a batch of ``tickets,'' which are subsequently used to cover the on-chain inclusion costs of transactions. When a user intends to initiate a transaction from a stealth address, they present the aggregator with one such ticket. This ticket is encoded using a Chaumian blinding scheme, a protocol widely employed in the privacy-focused e-cash systems first proposed in 1983~\cite{chaum1983}. Upon receipt of the ticket, the aggregator funds the recipient's account, bundles the transaction with others, and includes it within a block. Given that the funds involved in this process are minimal and exclusively used for transaction fees, the trust prerequisites are significantly lower than those associated with a full-scale implementation of privacy-preserving e-cash. This approach has significant potential to bridge the gap between privacy and functionality of stealth addresses.

\item[Stealth Address Detection. ]
A critical balance between privacy and detectability must be considered when addressing stealth address transactions. A public on-chain announcement is made whenever a stealth address transaction occurs, potentially enabling blockchain forensics to discern related transactions. To mitigate this issue, it is possible to broadcast the announcement through different channels than the stealth address transaction, breaking the link between the announcements and the actual transactions.

Detectability is not an issue exclusive to Ethereum-based privacy tools. Prominent Bitcoin CoinJoin wallets, such as Wasabi and Samurai Wallet, also integrate techniques enabling users to mix their funds and obscure their origins. Despite these efforts, the resulting transactions may still be identifiable using certain heuristic techniques~\cite{Ficsor_2017, Tironsakkul2022, Wahrstätter}. Similarly, Tornado Cash, a popular privacy tool within the Ethereum ecosystem, confronts a comparable challenge, as any deposit and withdrawal can be identified through publicly available event logs. However, it is crucial to recognize that external observers cannot de-anonymize the recipient of a stealth address interaction without access to the shared DH secret between the sender and the recipient.

\end{description}

\section{Conclusion}\label{sec:conclusion}
Stealth addresses have significant potential to enhance the privacy of programmable blockchain interactions. This work proposes BaseSAP as a blockchain-based foundation-layer protocol for stealth addresses compatible with different cryptographic schemes. 

BaseSAP is designed to function entirely autonomously, leveraging the immutable nature of Smart Contracts to deliver the required functionality for deploying interoperable stealth addresses on programmable blockchains. Compared to the previous solutions, the protocol's modularity not only encourages the evolution of cohesive auxiliary layers on top of its core implementation but also underscores its adaptability in accommodating various user applications, such as programmable wallets, public goods funding, Decentralized Finance (DeFi), Non-Fungible Tokens (NFTs), and more.

Through the simulations of an optimized Secp256k1-based stealth address protocol, we demonstrated the operational effectiveness of BaseSAP, the results of which we validated on the Goerli and Sepolia test networks via our initial prototype implementations.

Additionally, we conducted thorough cost analyses and identified possible security vulnerabilities and attack vectors that could undermine the privacy offered by stealth address protocols. The findings accentuate the necessity of confronting privacy issues prevalent in public and distributed blockchains.

In conclusion, our research provides the basis for implementing stealth address technology on programmable blockchains. It demonstrates the efficacy of BaseSAP in augmenting the privacy of public blockchain transactions. Furthermore, it underlines the significant potential of such protocols, mainly when applied on the application layer of programmable blockchains, to enhance interoperability across various aspects of blockchain technology. The code base created for this work is available under an open-source license to ensure reproducibility and transparency~\cite{nerolation2023}. The described protocol is available as an ERC (Ethereum Request for Comment)~\cite{eip5564}.




\ifCLASSOPTIONcaptionsoff
  \newpage
\fi




\bibliographystyle{IEEEtran}
%
\bibliography{main}



%








\begin{IEEEbiography}[{\includegraphics[width=1in,height=1.25in,clip,keepaspectratio]{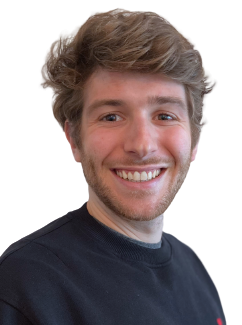}}]{Anton Wahrstätter}
received his BSc in Economics and LL.B. in Business Law from the University of Innsbruck, Austria, in 2018/2019. He completed his MSc in Digital Business from the University of Innsbruck, Austria, in 2020 and an MSc in Blockchain and Digital Currency from the University of Nicosia, Cyprus, in 2022. He has been contributing as an Open Source developer to the Bitcoin and Ethereum community since 2016, primarily focusing on privacy and quantitative data analysis. He is currently affiliated with the Institute for Distributed Ledgers and Token Economy at the Vienna University of Economics and Business, and the Research Institute for Cryptoeconomics in Vienna, Austria. His research focuses on blockchain privacy and trust, as well as the application of data science techniques to blockchain data.
\end{IEEEbiography}

\begin{IEEEbiography}[{\includegraphics[width=1in,height=1.25in,clip,keepaspectratio]{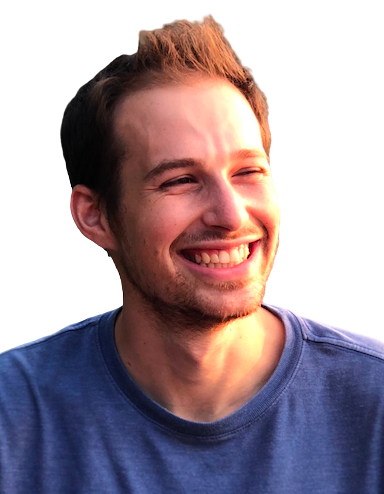}}]{Matthew Solomon}
received his Bachelor of Science (B.S.) in Aerospace Engineering from the University of Miami, US in 2014 and an M.S. in Aerospace, Aeronautical, and Astronautical/Space Engineering from the University of Maryland, US in 2016. He then spent several years at Lockheed Martin as an aerospace engineer and developed a keen interest in the field of cryptocurrencies. His academic and professional pursuits have led him to cultivate a deep interest in the intersections of open-source software development, privacy, and blockchain technology. Presently, Matt contributes to these domains, focusing on the design and implementation of privacy-oriented smart contract mechanisms.
\end{IEEEbiography}

\begin{IEEEbiography}
[{\includegraphics[width=1in,height=1.25in,clip,keepaspectratio]{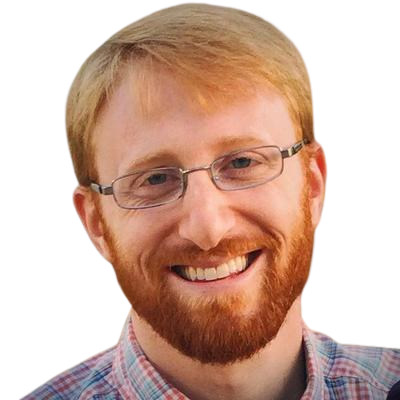}}]{Ben DiFrancesco}
earned his degree in Aerospace Engineering and began his career at Boeing as an aerospace engineer. In 2013, around the same time he discovered Bitcoin, he founded ScopeLift, a company initially focused on native mobile development. As his interest in cryptocurrencies grew, Ben pivoted the focus of ScopeLift towards the emerging crypto ecosystem. An engineer at heart, Ben has fostered a culture of technical excellence at ScopeLift. Currently, he is deeply involved in enhancing privacy on blockchains like Ethereum, particularly through the implementation of stealth address protocols.
\end{IEEEbiography}

\begin{IEEEbiography}
[{\includegraphics[width=1in,height=1.25in,clip,keepaspectratio]{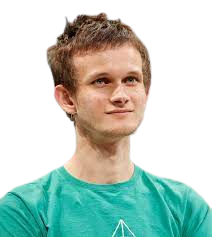}}]{Vitalik Buterin}
is a renowned computer scientist and programmer, best known for co-founding Ethereum, a pioneering platform in blockchain technology. His engagement in the world of cryptocurrency began in 2011 when he co-founded Bitcoin Magazine. His most significant contribution came in 2015 with the launch of Ethereum. Buterin has written numerous influential papers and is a highly-cited researcher in the field of blockchain technology. His research interests include blockchain technology, cryptoeconomics, consensus protocols, privacy and scalability solutions, and the security of blockchain systems and smart contracts. His pioneering work on Ethereum has significantly impacted the landscape of blockchain technology and the broader field of cryptocurrency.
\end{IEEEbiography}

\begin{IEEEbiography}[{\includegraphics[width=1in,height=1.25in,clip,keepaspectratio]{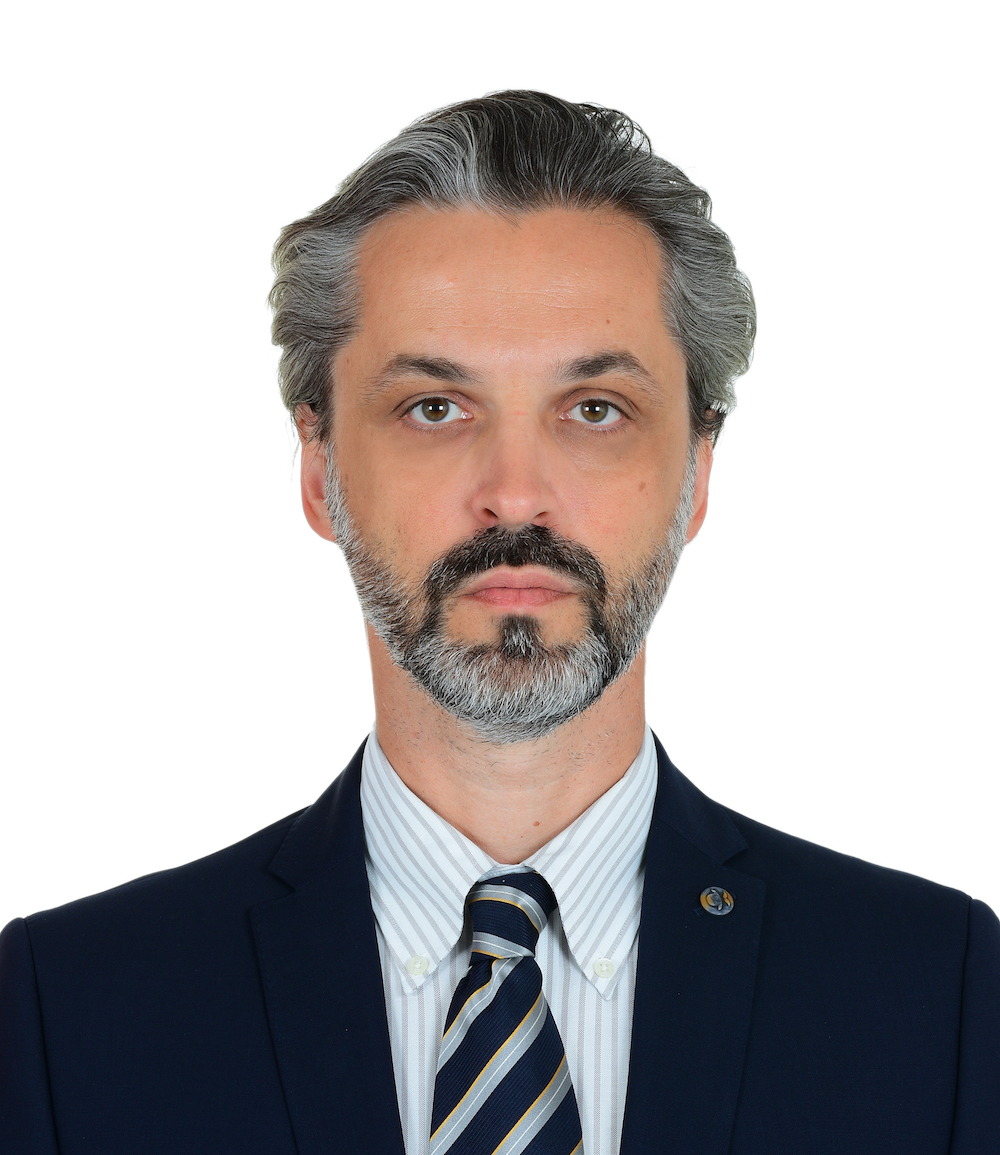}}]{Davor Svetinovic} (SM'16)
is a professor of computer science at the Department of Electrical Engineering and Computer Science, Khalifa University, Abu Dhabi, and the Department of Information Systems and Operations Management, Vienna University of Economics and Business, Austria (on leave), where he is the head of the Institute for Distributed Ledgers and Token Economy, and the Research Institute for Cryptoeconomics. He received his doctorate in computer science from the University of Waterloo, Waterloo, ON, Canada, in 2006. Previously, he worked at TU Wien, Austria, and Lero -- the Irish Software Engineering Center, Ireland. He was a visiting professor and a research affiliate at MIT and MIT Media Lab, MIT, USA. Davor has extensive experience working on complex multidisciplinary research projects. He has published more than 95 papers in leading journals and conferences and is a highly cited researcher in blockchain technology. His research interests include cybersecurity, blockchain technology, cryptoeconomics, trust, and software engineering. His career has furthered his interest and expertise in developing advanced research capabilities and institutions in emerging economies. He is a Senior Member of IEEE and ACM, and an affiliate of the Mohammed Bin Rashid Academy of Scientists.
\end{IEEEbiography}

\end{document}